# 5. Mentorship Network Structure: How Relationships Emerge Online and What They Mean for Amateur Creators

RUBY DAVIS, JENNA FRENS, NIHARIKA SHARMA, MEENA DEVII MURALIKUMAR, CECILIA ARAGON, AND SARAH EVANS

**Abstract:** Relationships form the core of connected learning. In this study, we apply and extend social network analysis methods to uncover the layered network structure of relationships among Fanfiction.net authors and reviewers. Fanfiction.net, one of the world's largest fanfiction communities, is a space where millions of young people engage with written media, connect over shared interests, and receive support and mentoring from a distributed audience. Does an affinity space such as Fanfiction.net have the same structure as social networks Facebook and Twitter? We applied k-means clustering on millions of relationships to determine that Fanfiction.net has 2 to 3 layers, in contrast with the 4-layer structure of Facebook and 5-layer structure of Twitter. In addition, we conducted a large-scale machine classification of Fanfiction.net reviews to reveal the types of mentoring exchanged in each layer. Our findings show that the relationships where reviews are exchanged most frequently are most likely to contain substantive reviews. We discuss implications of these findings for the theory of distributed mentoring as well as the design of online affinity networks.

## Introduction

Relationships in online affinity spaces provide emotional support, connection to employment, and learning opportunities to a generation of digitally connected young people (Ito et al., 2019). One such space, Fanfiction.net, hosts more than 10 million stories written by authors primarily aged 13 to 26. Between 2001 and 2018, 53 million relationships emerged among Fanfiction.net users. Quantitatively examining the network structure of Fanfiction.net can provide a view into how relationships emerge and their role in informal learning.

In this study, we extended Dunbar's social network analysis method (Dunbar, Arnaboldi, Conti, & Passarella, 2015) to reveal the two- to three-layer social structure of distributed mentoring on Fanfiction.net. We classified reviews according to category within a distributed mentoring framework and characterized relationships according to the type of support most often exchanged. Our analysis centers on the following two research questions:

1. How many layers are in a Fanfiction.net user's ego network structure?
2. What kinds of reviews are exchanged in each layer?

Prior work has shown that social networks Facebook and Twitter mirror the layered network structure of the offline world (Dunbar et al., 2015). Our work generalizes this theory to an affinity network and reveals differences between socially driven and interest-driven spaces. We also extend the theory of distributed mentoring (Campbell et al., 2016) by quantitatively examining the function of different relationships within social layers. This knowledge can aid in the construction of more effective mentorship communities whose affordances encourage and enhance distributed mentoring structures.

# Background

## Fanfiction and Affinity Networks

*Fanfiction*, or written works produced by fans who use characters or settings from preexisting media, plays a critical role in a vast digital participatory culture (Jenkins, 1992) that builds relationships around sharing, creating, and consuming content. One such community that centers around a given piece of media is a *fandom*. Fanfiction provides a vital outlet for marginalized people to express themselves, allowing those who are traditionally excluded from mainstream media production to tell their own stories (Aragon & Davis, 2019; Black, 2008). Today's fanfiction is primarily published and distributed online within fanfiction communities and archives, such as Fanfiction.net and Archive of Our Own. These sites are particularly popular among young people and represent part of a greater movement of youth participation in online digital media.

Fanfiction communities also serve as *affinity spaces*, or places where individuals of all skill levels may come together over shared interests (such as a TV show or book) to learn from one another (Gee, 2004; Ito et al., 2019). Affinity spaces such as online fanfiction communities exist within the greater context of shared culture, creation, skill building, and connection with peers known as an *affinity network* (Ito et al., 2019). As interest-driven online networks, affinity networks are distinct from many other popular social networks such as Facebook or Twitter, where the social network structure is already known. Our research unveils the previously unknown network structure of an affinity network and extends our understanding of affinity networks by examining where in the network different types of relationship behaviors may occur.

## Distributed Mentoring

Understanding the structure of relationships in fanfiction communities can extend the theory of distributed mentoring and provide further insight into how it may be fostered. *Distributed mentoring* describes a network-based collaborative mentoring process in affinity spaces (such as Fanfiction.net) built upon computer-mediated interactions (Aragon & Davis, 2019; Campbell et al., 2016; Evans et al., 2017). In distributed mentoring relationships, the roles of mentor and mentee are fluid; an author may mentor others in areas such as grammar while receiving feedback on world building. The feedback exchanged in distributed mentoring relationships within fanfiction communities can take a variety of forms, including *targeted constructive*, or reviews that offer constructive criticism; *targeted positive*, or reviews positively reflecting on specific aspects of the story; and *update encouragement*, or reviews that encourage further writing (Evans et al., 2017).

The theory of distributed mentoring has characterized these mentoring relationships and categorized the kinds of conversations that occur within them; however, little is known about the structure of the relationships in distributed mentoring networks. In addition, connections have yet to be drawn between the content of an exchange between individuals and where in an individual's network the exchange occurs. As relationships are the backbone of distributed mentoring, the context and history of a given distributed mentoring relationship may completely transform the meaning and impact of distributed mentoring artifacts, such as comments or reviews. Reviews from ongoing relationships, for example, may be received and interpreted very differently from a single one-off review. A model of distributed mentoring that incorporates this context will better account for the influence of informal learning in online spaces. To build this model, we need to quantify relationships and qualify the exchanges that occur within them.

## Dunbar's Circles

The number of relationships a person can maintain is finite. These relationships may be categorized into several distinct layers of closeness termed *Dunbar's circles* (Dunbar et al., 2015). The smallest circle consists of about five people (or alters) with whom the individual (or ego) has a very close relationship. Every subsequent circle increases in size and decreases in closeness until the cognitive limit is reached. The theory of Dunbar's circles holds for online relationships; on Twitter and Facebook, the structure of one's virtual relationships is similar to that observed in face-to-face relationships (Dunbar et al., 2015).

Our research explores whether such a hierarchy exists in distributed mentoring networks. We conduct Dunbar's analysis over a network of fanfiction authors and reviewers to uncover the number of mentees and frequency of mentoring in each distinct circle. We then we extend this method further by characterizing each layer.

## **Method**

We determine the number of layers in the social networks of Fanfiction.net's active authors and reviewers. Then we quantify each layer's review types as classified by a machine learning algorithm.

## The Fanfiction.net Data Set

We collected a data set from Fanfiction.net that contains the stories, author profiles, reviews, and associated metadata of activity on the site from January 2001 to January 2017. This massive data set contains approximately 28 million chapters of fanfiction and 177 million reviews written by an international group of 10 million young people. Previous literature further describes the collection and demographics of this data set (Bathija & Tekriwal, 2019; Frens, Davis, Lee, Zhang, & Aragon, 2018; Yin, Aragon, Evans, & Davis, 2017). Each review in the data set is associated with a time stamp from when it was posted to Fanfiction.net and IDs of the reviewer and author in the exchange. Using this data, we built a social graph of relationships on Fanfiction.net.

For the purpose of our analysis, we defined relationships as reviewer-author pairs with at least two reviews exchanged over a duration of at least one month. We define the contact frequency of a relationship as the number of times that social contact occurred over the duration of the relationship. To compute contact frequency, we divided the number of reviews by the amount of time, in months, between the first and last review. If the reviewer has written stories of his or her own, and the author reviews them, we represent this kind of relationship with bidirectional edges from reviewer to author and vice versa. In this graph, each user, or ego, is a node. Every reviewer-author pair is a directional edge, weighted by interaction frequency in reviews per month. Anonymous reviews were excluded from this analysis. The resulting graph contains 53,202,307 relationships between 2,580,411 reviewers and 1,373,910 authors.

## Dunbar's Approach

We used Dunbar et al.'s k-means clustering approach to determine the optimal number of layers in a social network (Dunbar et al., 2015). As in Dunbar's study, we analyzed the subset of users who are active in the community, either giving or receiving an average of 10 reviews per month or more. In addition to this restriction, we included only users with at least 25 connections. As our data set contains two different kinds of participants, reviewers and authors, we performed two versions of the analysis: one with reviewers as egos and one with authors as egos. Each involves a different data set of reviewers and authors based on the directional application of the inclusion criteria.

After we applied the inclusion criteria, the reviewer-as-ego data set contained 62,869 reviewers (egos) and 458,286 authors (alters), or 2.4% of the reviewer population in the Fanfiction.net data set. These reviewers reviewed 59.4 authors on average with a standard deviation of 65.6. The average interaction frequency of a relationship (edge) in this data set was 1.86 reviews per month ($SD$ = 2.48), and there were 3,733,123 relationships in total. The reviewer-as-ego data set contained 31.4% of all reviews on Fanfiction.net.

The author-as-ego data set contained 66,798 authors (egos) and 747,502 reviewers (alters), or 4.86% of the author population in the Fanfiction.net data set. The authors in the data set received reviews from 70.87 reviewers on average with a standard deviation of 96.43. The average interaction frequency of a relationship (edge) was 1.85 reviews per month ($SD$ = 2.52), and there were a total of 4,734,203 relationships. The author-as-ego data set contained 73.03% of all reviews exchanged on Fanfiction.net.

We extracted ego networks from the data set by aggregating edges per ego. We performed k-means clustering on each ego, varying k from 1 to 20. The k-means algorithm partitions an ego's edges, based on their weight, into k separate clusters with minimum sum of squared Euclidean distance between edges and cluster centers (MacQueen, 1967). As in Dunbar's analysis, we computed x, the optimal k for each ego in the data set. We then chose the overall optimal number of clusters k* using the elbow method, taking the value of k that most efficiently accounts for variance in Euclidean distance (Kodinariya & Makwana, 2013). We computed p(x), the proportion of ego networks where k* = x, that is, they are optimally clustered with x clusters. Figure 1 shows the probability distribution of p(x) for reviewer-as-ego and author-as-ego data sets. We calculated s(x), the silhouette score, to determine how closely the data are matched to data within their cluster (Rousseeuw, 1987). Values of s(x) close to 1 show that the data are optimally clustered.

## Categorizing Reviews Exchanged in Each Network Layer

We categorized the reviews in the Fanfiction.net data set by using a qualitatively coded subset of reviews as ground truth for a machine classifier. The grounded theory approach and qualitative coding process are described in prior work (Evans et al., 2017). A group of researchers coded 3,566 reviews by category within the distributed mentoring framework and combined this with the original qualitatively coded data set (Evans et al., 2017) for a total of 8,066 reviews. For the present analysis, we focused on three nonexclusive review categories: update encouragement, targeted positive reviews, and targeted constructive. We combined the latter two categories into a single "targeted" category to differentiate substantive reviews from update encouragement.

We then trained a machine classifier on the human-coded data set to classify all reviews in the Fanfiction.net data set. We improved and validated this method of qualitative to quantitative classification as described in detail by Scott et al. (2012), and the tools used in this process are outlined in detail by Michael Brooks (2015). Using the tool ALOE, we classified the reviews in the Fanfiction.net data set with 87% accuracy for update encouragement and 75% accuracy for targeted reviews. The review classification is further described in our blog (Bathija & Tekriwal, 2019). We then used this classification to examine the proportion of each type of review exchanged in each network layer.

# Findings

## The Two- to Three-Layer Network Structure of Fanfiction Reviewing

In the reviewer-as-ego data set, we determined the optimal number of clusters to be k* = 2, 3 using the elbow method, indicating that active Fanfiction.net reviewers have a two- to three-layer network. During the k-means analysis described above, we found the mean optimal number of clusters to be x = 2.39. Visualizing p(x) in Figure 1, we see a clear drop-off from 2 to 3 and from 3 to 4, and subsequent flattening of the graph. The mean silhouette score of ego networks clustered was s(x) = 0.6952, indicating that this is an appropriate and optimal clustering of the data set and the identified clusters are factual. In the author-as-ego data set we also confirmed that active Fanfiction.net authors have a two- to three-layer network. The mean optimal number of clusters was x = 2.36. Visualizing p(x) in Figure 1, we found a similar distribution to the reviewer-as-ego data set and determined k* = {2, 3}, with s(x) = 0.6857.

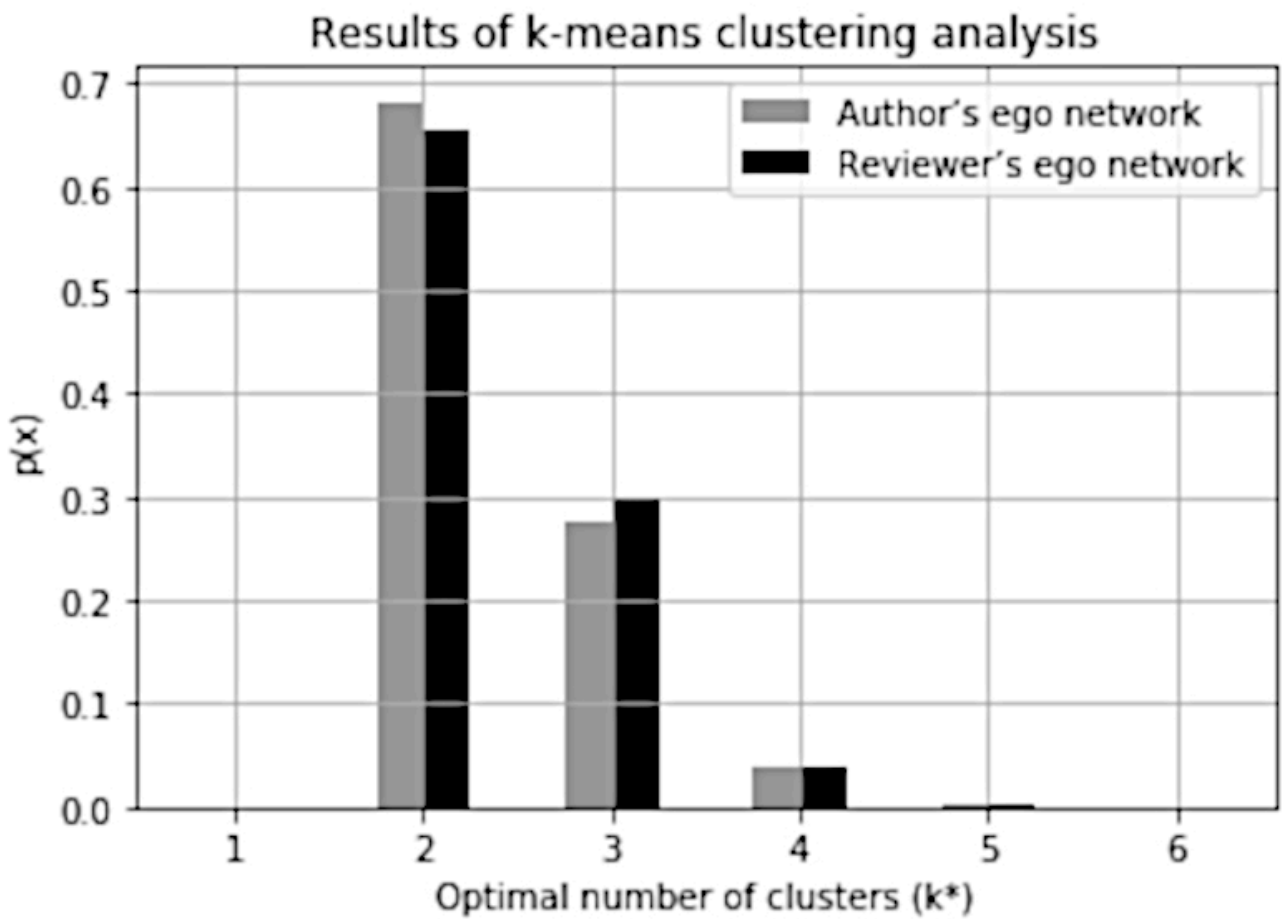

Figure 1. The probability that an ego network is optimally clustered for (a) author's ego network, (b) reviewer's ego network, p(x), for x = k with k ranging from 1 to 6 network layers (we experimented with k ranging from 1 to 20 network layers, but layers after 6 have very low p(x) values).

Figure 1 demonstrates a marked peak at k = 2 and k = 3 for both reviewer's ego and author's ego network data set, with an average optimal number of cluster size ranging between 2.36 to 2.38 for both data sets. To study the size and mean contact frequency of the ego network layer we used k = {2, 3}, which encapsulates both the median and mean of each data set. We summarize the size and mean monthly contact frequency of each layer in Table 1 and interpret these results in the next two sections.

|  | Layer | 0 | 1 | 2 |
|---|---|---|---|---|
| Reviewer as ego (k=2) | Number of Alters | 9.08 ± 9.10 | 50.30 ± 59.34 | - |
|  | Contact Frequency | 7.01 ± 7.50 | 1.26 ± 0.65 | - |
| Reviewer as ego (k=3) | Number of Alters | 3.89 ± 3.52 | 14.76 ± 13.98 | 40.73 ± 51.82 |
|  | Contact Frequency | 8.66 ± 8.04 | 3.25 ± 2.92 | 0.97 ± 0.58 |
| Author as ego (k=2) | Number of Alters | 11.48 ± 12.39 | 59.40 ± 89.79 | - |
|  | Contact Frequency | 6.77 ± 8.36 | 1.26 ± 0.72 | - |
| Author as ego (k=3) | Number of Alters | 4.72 ± 4.57 | 17.997 ± 19.45 | 48.16 ± 78.86 |
|  | Contact Frequency | 8.38 ± 8.82 | 3.11 ± 3.32 | 1.02 ± 0.70 |

*Table 1. Mean (±SD) number of alters and contact frequency in each layer identified by k-means analysis.*

In the reviewer-as-ego data set, a reviewer's ego network is likely to have two or three layers. The innermost layer, layer 0, has a mean of 9.08 alters (for two-layer networks) or 3.89 alters (for three-layer networks). Reviewers review the authors in this layer around seven to eight times a month on average. This indicates that reviewers maintain a close reviewing relationship with a small group of authors, with individual variation causing contact to range from every day to a few times per month. The outermost layer has a mean of 50.30 alters (for two-layer networks) or 40.73 alters (for three-layer networks) and a mean frequency of about one review per month. This indicates that reviewers infrequently review a large group of authors that ranges widely in size. Three-layer networks also have a middle layer, which has a mean of 14.76 alters with a mean contact frequency of 3.25 reviews per month.

The author-as-ego data set revealed that authors are also likely to have ego networks with two or three layers. The author ego's closest circle, layer 0, has a mean of 11.48 alters (for two-layer networks) or 4.72 alters (for three-layer networks). Authors receive about seven or eight reviews per month from reviewers in this layer. This indicates that authors experience close but not necessarily daily contact from a small group of reviewers, with individual variation causing contact to range from daily to almost weekly. The outermost layer has a mean of 59.40 alters (for two-layer networks) or 48.16 alters (for three-layer networks). Authors get reviews about once a month from reviewers in this layer. This indicates authors are likely to receive infrequent contact from a large following that ranges widely in size. Three-layer networks also have a middle layer with a mean of 17.997 alters and a mean contact frequency of 3.11 reviews per month. We visualize an example author ego network in Figure 2.

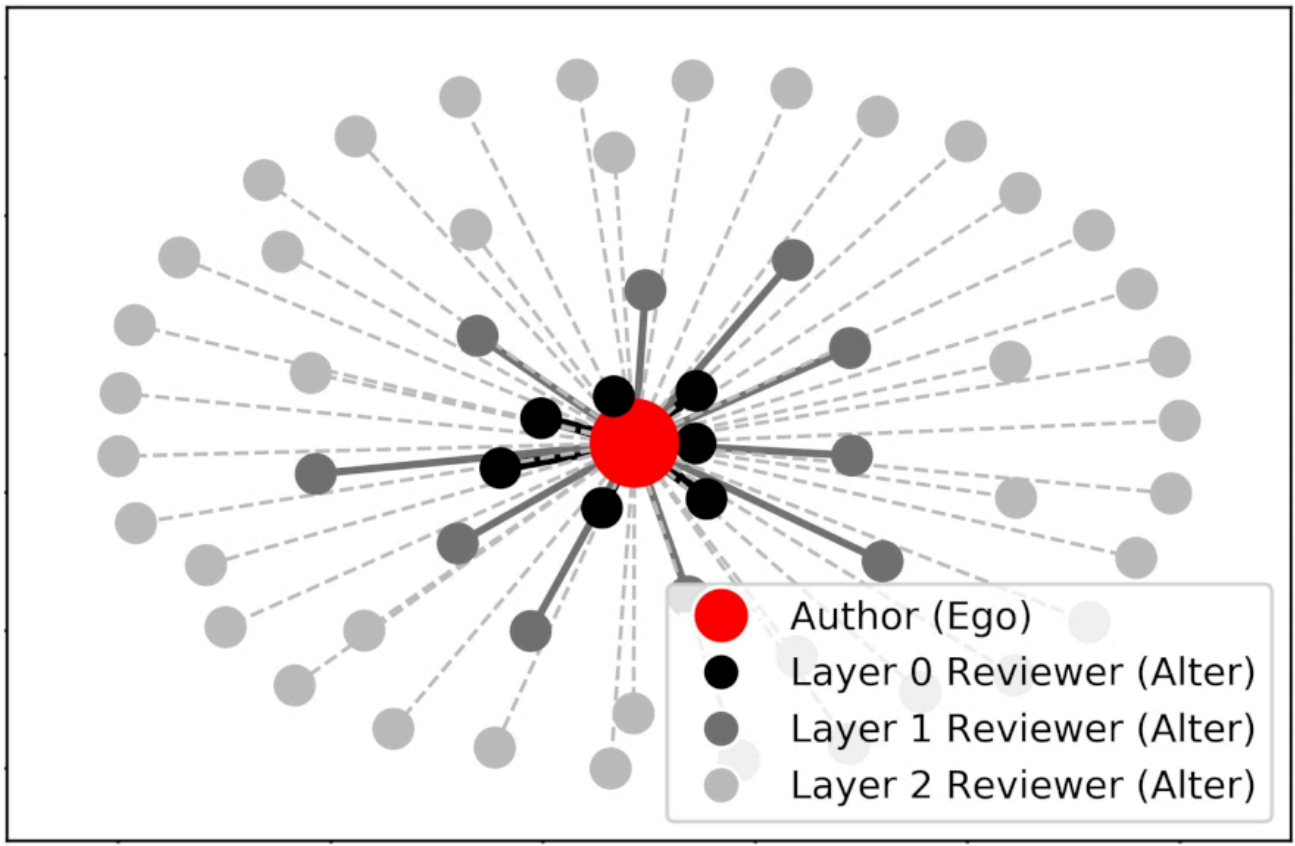

*Figure 2. Visualization of an author's relationships, colored by layer and weighted by contact frequency.*

Figure 2 depicts a Fanfiction.net author's ego network. For the particular author in Figure 2 (red node), we can see the connected reviewers (gray and black nodes) and the corresponding layers they fit in. The black nodes signify the closest layer, the dark gray nodes signify the middle layer, and the light gray nodes signify the final layer. This graph is weighted according to the frequency of contact for a given author-reviewer pair.

## Types of Reviews Exchanged by Layer

We computed a count of update encouragement and targeted reviews in each layer of the reviewer-as-ego data set. Update encouragement reviews contain statements to the author that he or she should continue publishing work. Targeted reviews contain substantive commentary about the writing. These categories are not exclusive. The counts for each layer are shown in Table 2.

|  | Layer | 0 | 1 | 2 |
|---|---|---|---|---|
| Count (Percentage) of Reviews | Update Encouragement | 440,626 (17.6%) | 3,179,649 (24.3%) | 7,927,321 (27.9%) |
|  | Targeted | 1,283,701 (55.3%) | 6,948,617 (53.1%) | 14,161,050 (49.8%) |

*Table 2. Count of each review type in each of the layers of a reviewer's network structure.*

As shown in Table 2, the proportion of update encouragement increases from layer 0 to layer 2, while the proportion of targeted reviews decreases. This indicates that reviewers spend more effort on substantive reviews in relationships with frequent contact. We discuss the implications of these findings below.

**Discussion**

Our analysis revealed that active authors and reviewers tended to maintain a small number of close relationships (4 to 12) with both high frequency of contact (seven to nine times a month) and higher likelihood of exchanging effortful and substantive reviews (55.7%). Authors and reviewers also had a larger number of relationships (40 to 60) that were infrequent (one to three times a month) and more likely to contain less effortful reviews such as update encouragement (27.9%). These findings suggest that fanfiction participants tend to saturate their mentoring networks up to similar cognitive limits theorized by Dunbar et al. for more generalized social networks (Dunbar et al., 2015). We contextualize these findings within prior literature and discuss implications for the design of online affinity spaces.

## Comparison With Facebook and Twitter

Although the social network structures are similar, there are fewer layers in the ego networks of active Fanfiction.net users in comparison with Facebook and Twitter users (two and three versus four and five respectively [Dunbar et al., 2015]). The closest layer seen on Twitter, with alters contacting approximately at least once every one to two days, did not exist on Fanfiction.net. This may be explained by the affordances of reviewing fanfiction: Reviewer-author relationships are reliant on an author's story update schedule, and readers are unlikely to leave reviews without the catalyst of a new chapter's being published. Twitter users may simply be able to create content more frequently because tweets are typically shorter than fanfiction chapters, resulting in more frequent exchanges. Additionally, Fanfiction.net lacked the middle layer seen on Facebook and Twitter, showcasing the two types of reviewers on the platform: frequent or infrequent. The outermost layer of the Fanfiction.net data set and the Facebook and Twitter data set demonstrated similar behavior with some active reviewers typically maintaining low-contact relationships (the frequency of the outermost layer on Fanfiction.net was 1.26/month; on Facebook it was 1.37 and on Twitter 2.54).

## Comparison With Previous Findings

These results suggest a difference between interest-driven and socially driven online communities, or "hanging out" versus "geeking out" (Ito et al., 2019). Fans behave in a "nomadic" fashion, hopping from source material to source material as their interests change (Jenkins, 1992). Throughout the course of a reviewer-author relationship, a reviewer may lose interest in the story, unsubscribe from the author, or migrate to a new fandom. We also found high individual variation; the ranges in number of alters per layer were larger than Facebook and Twitter. This indicates a diversity of behavior among active reviewers on Fanfiction.net. Some of this variation may be a result of differences between fandoms–particularly small and large fandoms. Interest and participation in a smaller fandom may lead to fewer and closer connections than in a large fandom (Aragon & Davis, 2019).

The theory of distributed mentoring (Campbell et al., 2016) provides a framework for understanding the value and contribution of different feedback relationships. A large outer circle of reviewers helps to provide an abundance of feedback in the form of shallow positive reviews and update encouragement, which in turn are linked to continued participation (Bathija & Tekriwal, 2019) and language development (Frens et al., 2018). The inner circle provides specific, directive feedback, which authors value highly and reciprocate, leading to stronger relationships, shared context, and better feedback in a virtuous cycle. The closer the relationship, the more effort is invested in feedback exchange, oftentimes going beyond the reviewing platform to beta reading, in-progress feedback, and ideation feedback (Aragon & Davis, 2019).

## Design Implications

Our findings imply that fanfiction websites and other online affinity spaces could be designed to optimize distributed mentoring by considering Dunbar's theory in the implementation of affordances for connection, feedback, exposure, and recommendation. For instance, Fanfiction.net and Archive of Our Own both default to sorting works by their publication date—the most recently updated fanfiction works are most likely to be seen. The result is that authors who publish frequently are most likely to receive exposure. However, based on our findings, these authors are less likely to need additional exposure in comparison with new authors who are posting for the first time in that fandom. There is potential for designing a different default that grants new authors more exposure until their networks are close to saturation. Likewise, for users with saturated networks, designs could focus on deepening their current relationships over exposure to more connections. Furthermore, we find that readers who frequently review an author are more likely to give substantive feedback. Platforms could further encourage this behavior by reminding frequent readers to do so, while prompting less frequent readers to give low-effort reviews such as update encouragement.

## Future Research

Our research focused only on substantive reviews and update encouragement, but further research could examine the occurrence of additional review types in a reviewer-author network. Examining how the bidirectionality of communication may differ between layers could also bolster our understanding of what each layer represents. Furthermore, the reviews that are exchanged on Fanfiction.net represent a small portion of the many interactions that occur in fandoms across the net. Examining how fans engage with one another across many different platforms could provide a more holistic view of this affinity network and its structure. This research could also be extended to explore networks beyond fan communities by comparing Fanfiction.net's structure to that of other networks, such as in-person mentorship networks.

## **Conclusion**

Our large-scale analysis revealed the relationship structures within a distributed mentoring network. We characterized different layers of reviewer and author relationships and showcased the number, size, and distribution of relationship layers on Fanfiction.net. The findings show that targeted feedback is most likely to occur in the innermost layers, while less targeted feedback occurs more frequently in the outermost layers. To facilitate an ideal environment for distributed mentoring, sites must encourage the development of close relationships, where targeted feedback is exchanged, without oversaturating an individual's network beyond the limited number of relationships he or she can maintain. Future research can explore this design space to ensure the optimal balance of exposure and connection that allows individuals to build networks that facilitate mentoring. Designs that actively foster distributed mentoring networks can help millions of young people build their writing skills and share their voices.

# Acknowledgments

We thank the passionate fanfiction community for sharing their stories, reviews, and perspectives.